\documentclass{article}

\usepackage{cuted}
\usepackage{microtype}
\usepackage{graphicx}
\usepackage{subfigure}
\usepackage{booktabs} 
\usepackage{tikz}
\usepackage{listings}
\usepackage{float}

\lstnewenvironment{algolst}[1][]
{
  
  \lstset{#1}
}{}

\usepackage{hyperref}

 \usepackage[accepted]{icml2023}

\usepackage{amsmath}
\usepackage{amssymb}
\usepackage{mathtools}
\usepackage{amsthm}
\usepackage{physics}

\usepackage[capitalize,noabbrev]{cleveref}

\theoremstyle{plain}

\theoremstyle{definition}

\theoremstyle{remark}

\icmltitlerunning{Time Delay Cosmography with a Neural Ratio Estimator}

\begin{document}

\twocolumn[
\icmltitle{Time Delay Cosmography with a Neural Ratio Estimator}




\begin{icmlauthorlist}
\icmlauthor{\`{E}ve Campeau-Poirier}{udem,ciela,mila}
\icmlauthor{Laurence Perreault-Levasseur}{udem,ciela,mila,flatiron,perimeter}
\icmlauthor{Adam Coogan}{udem,ciela,mila}
\icmlauthor{Yashar Hezaveh}{udem,ciela,mila,flatiron,perimeter}
\end{icmlauthorlist}

\icmlaffiliation{udem}{Department of Physics, Université de Montréal, Montréal, Canada}
\icmlaffiliation{mila}{Mila, Montréal, Canada}
\icmlaffiliation{ciela}{Ciela, Montréal, Canada}
\icmlaffiliation{flatiron}{Flatiron Institue, New York, USA}
\icmlaffiliation{perimeter}{Perimeter Institute for Theoretical Physics, Waterloo, Ontario, Canada}

\icmlcorrespondingauthor{\`{E}ve Campeau-Poirier}{eve.campeau-poirier@umontreal.ca}

\icmlkeywords{Hubble constant ---
        Strong gravitational lensing ---
        Neural ratio estimator ---
        Simulation-based inference ---
        Machine learning}

\vskip 0.3in
]



\printAffiliationsAndNotice{}  

\begin{abstract}
We explore the use of a Neural Ratio Estimator (NRE) to determine the Hubble constant ($H_0$) in the context of time delay cosmography. Assuming a Singular Isothermal Ellipsoid (SIE) mass profile for the deflector, we simulate time delay measurements, image position measurements, and modeled lensing parameters. We train the NRE to output the posterior distribution of $H_0$ given the time delay measurements, the relative Fermat potentials (calculated from the modeled parameters and the measured image positions), the deflector redshift, and the source redshift. We compare the accuracy and precision of the NRE with traditional explicit likelihood methods in the limit where the latter is tractable and reliable, using Gaussian noise to emulate measurement uncertainties in the input parameters. The NRE posteriors track the ones from the conventional method and, while they show a slight tendency to overestimate uncertainties, they can be combined in a population inference without bias.

\end{abstract}


\section{Introduction}
\label{sec:intro}

Over the past decades, the inflationary $\Lambda$CDM model has had striking success in explaining cosmic microwave background (CMB) observations and the detailed evolution of the Universe. The current expansion rate of the Universe, known as the Hubble constant ($H_0$), is essential for many studies, including understanding the nature of dark energy, neutrino physics, and testing general relativity. In the past decade, the measured values of $H_0$ from different probes have diverged: the latest CMB and Type Ia supernovae data now disagree at more than 4$\sigma$ \citep{Riess2022}.



Time delay cosmography can provide an independent measurement of $H_0$ with different systematics from existing methods. This can be done using the time delays between the multiple images of a strongly lensed variable light source. Previous measurements have achieved a precision between 2\% and 8\% \citep{B&T2021} using this method. Meanwhile, 1\% precision is required to solve the Hubble tension \citep{Weinberg2013, Treu2022}. This could be achieved with data available in the next decade with a new generation of survey telescopes. The Rubin Observatory, in particular, is expected to detect thousands of strongly lensed quasars \citep{O&M2010}.

However, current analysis methods have limitations in terms of complexity and scalability. They rely on likelihood-based approaches, such as Markov Chain Monte Carlo (MCMC) and nested sampling, which require explicit likelihoods and are not amortized. They also require sampling joint posterior distributions of nuisance parameters while only the $H_0$ marginal is of interest. Hence, they scale poorly as nuisance parameters are included to ensure unbiasied inference.

The simulation-based inference (SBI) framework allows handling complex, high-dimensional data and models that are difficult or intractable to analyze using traditional likelihood-based methods by only relying on the availability of a realistic simulation pipeline. Neural Ratio Estimators (NREs; \citealt{Cranmer2015}), a specific class of SBI methods, leverage the power of machine learning to allow amortization of the inference process as well as implicit marginalization over large sets of nuisance parameters, providing an efficient way to estimate low-dimensional variables. 

We demonstrate the application of an NRE to time delay cosmography by predicting the $H_0$ posterior distribution given Fermat potentials calculated from modeled lens parameters and image positions, the time delay measurements, and the deflector and source redshifts. We use a Set Transformer architecture \citep{Lee2019}, which allows for the amortization over lensing systems with two or four lensed images by the same model.

While previous works have explored how machine learning can be used for the measurement of $H_0$ with time-delay cosmography, contributions (e.g. \citealt{Hezaveh2017, PerreaultLevasseur2017, Morningstar2019, Person2019,Wagner-Carena2021, Schuldt2021, Park2021}) have been limited to using neural networks (NN) to estimate the lens parameters posterior. The approach presented here is therefore complementary, since it bridges the remaining gap to fully amortize the inference of $H_0$ from strong lensing data.  

Section \ref{sec:cosmography} introduces the methodology. Section \ref{sec:simulations} describes the simulations. Section \ref{sec:NRE} presents the NN architecture and training procedure. Results are presented in section \ref{sec:results}.


\section{Time-delay cosmography}
\label{sec:cosmography}
Gravitational lensing occurs when images from a distant source get distorted by the presence of matter bending space-time along the line of sight. In strong gravitational lensing, there is formation of multiple images of background sources due to this effect. The lensing equation,
\begin{equation}
    \boldsymbol{\beta} = \boldsymbol{\theta} - \boldsymbol{\alpha}\left(\boldsymbol{\theta}\right) \, ,
    \label{eq:lensing}
\end{equation}
summarizes this phenomenon by retracing the source plane angular position $\boldsymbol{\beta}$ of a ray observed at the image plane angular position $\boldsymbol{\theta}$ after a mass deflector has deviated it by an angle $\boldsymbol{\alpha}$. The lensing potential $\psi$ of the massive object determines the angular deflection $\boldsymbol{\alpha}$ and the convergence $\kappa$ according to
\begin{equation}
    \boldsymbol{\alpha}\left(\boldsymbol{\theta}\right) = \boldsymbol{\nabla} \psi\left(\boldsymbol{\theta}\right) \, ;\quad \nabla^2 \psi \left(\boldsymbol{\theta}\right) = 2 \kappa\left(\boldsymbol{\theta}\right) \,.
    \label{eq:alpha}
\end{equation}

Gravitational lensing affects the light rays travel time from their source to the observer in two ways : by changing their path length and through the lensing potential itself. The presence of a mass deflector in the light's trajectory lengthens its travel time by an amount proportional to the Fermat potential $\phi$, which is fully determined by the mass distribution in the lens and is given by 
\begin{equation}
    \phi \left(\boldsymbol{\theta}, \boldsymbol{\beta} \right) \equiv \frac{\left(\boldsymbol{\theta} - \boldsymbol{\beta} \right)^2}{2} - \psi \left(\boldsymbol{\theta}\right) \, .
    \label{eq:Fermat}
\end{equation}

To infer $H_0$ with time delay cosmography, one observes a multiply-imaged time-varying background source. Each path giving rise to each image is affected by a different Fermat potential, resulting in a different light travel time. This allows the evaluation of the relative travel times between paths $\Delta t$, which are called time delays. They are calculated between pairs of images. They are related to $H_0$ by
\begin{equation}
    \Delta t \equiv \frac{D_{\Delta t}}{c} \Delta\phi \, ,
    \label{eq:time_delays}
\end{equation}
where $c$ is the speed of light, $\Delta \phi$ is the difference of Fermat potential at the position of the two distinct images, and $D_{\Delta t}$ is the time delay distance, given by
\begin{equation}
    D_{\Delta t} \equiv \left(1+z_d\right)\frac{D_d D_s}{D_{ds}} \, .
    \label{eq:D_dt}
\end{equation}
Here, $z_d$ is the deflector redshift, $D_d$ is the diameter angular distance between the observer and the deflector, $D_s$ is the diameter angular distance between the observer and the source, $D_{ds}$ is the diameter angular distance between the deflector and the source. These distances are where the $H_0$ dependence is contained.

In this framework, the posterior distribution of $H_0$ generally takes the form
\begin{equation}
    P(H_0 | \boldsymbol{\Delta t}, \boldsymbol{d}) \propto \int \, \mathrm{d}\boldsymbol{\zeta} \, P(\boldsymbol{\Delta t} | H_0, \boldsymbol{\zeta}, \boldsymbol{M}) P(\boldsymbol{\zeta} | \boldsymbol{d}, \boldsymbol{M}) P(H_0)
    \label{eq:general_posterior}
\end{equation}
where $\boldsymbol{d}$ represents the lensing observation, $\boldsymbol{\zeta}$ is a set of parameters describing the lensing system, and $\boldsymbol{M}$ includes all observational effects (e.g. instrumental noise, point spread function, image covariance matrix, deflector's light, and dust). In this context, the lensing parameters and the observational effects are nuisance parameters that must be integrated out to obtain the marginal distribution of $H_0$. The main proposal of this work is to replace the traditional Monte Carlo methods to numerically approximate the $H_0$ posterior. 


\section{Simulations}
\label{sec:simulations}

In this work, we consider the case where the deflected light is emitted by a variable point source, such as an Active Galactic Nucleus (AGN) or a supernova. We do not consider any light profile for its host galaxy because in the following we assume that the modeling of the lensed image was performed in a previous analysis stage (e.g. with a BNN as in \citealt{Park2021}).
 We assume that the source is being distorted by a deflector following as Singular Isothermal Ellipsoid (SIE; \citealt{Kormann1994}), plus external shear. This model is described by 7 parameters: Einstein radius $\theta_E$, $x-$ and $y-$components of the position $(x_d,\, y_d)$, axis ratio $f$ and its orientation $\phi_d$, and modulus $\gamma_\text{ext}$ and orientation $\phi_\text{ext}$ of the external shear. Details about the range of uniform prior used for these parameters, the cosmology, and the variable source are included in Table~\ref{tab:param}.

\begin{table}[t]
    \begin{center}
    \caption{Prior distributions of all the parameters needed to generate Fermat potentials and time delays in our framework}
    \label{tab:param}
    \begin{tabular}{lr}
    \toprule
    \textbf{Parameter} & \textbf{Distribution} \\
    \midrule
    \midrule
    \textbf{Cosmology} & \\
    \midrule
    Hubble constant (km s$^{-1}$ Mpc$^{-1}$) & $H_0 \sim \mathrm{U}(40, 100)$\\
    Dark energy density & $\Omega_\Lambda = 0.7$ \\
    Matter energy density & $\Omega_m = 0.3$ \\
    \midrule
    \textbf{Deflector} & \\
    \midrule
    Redshift & $z_d \sim \mathrm{U}(0.04, 0.5)$ \\
    Position ($''$) & $x_d, y_d \sim \mathrm{U}(-0.8, 0.8)$ \\
    Einstein radius ($''$) & $\theta_E \sim \mathrm{U}(0.5, 2.0)$\\
    Axis ratio & $f \sim \mathrm{U}(0.30, 0.99)$ \\
    Orientation (rad) & $\varphi_d \sim \mathrm{U}(-{}^\pi/_2, {}^\pi/_2)$ \\
    \midrule
    \textbf{External Shear} & \\
    \midrule
    Modulus & $\gamma_\text{ext} \sim \mathrm{U}(0, 0.2)$ \\
    Orientation (rad) & $\varphi_\text{ext} \sim \mathrm{U}(-{}^\pi/_2, {}^\pi/_2)$ \\
    \midrule
    \textbf{Variable point light source} & \\
    \midrule
    Redshift & $z_s \sim \mathrm{U}(1, 3)$ \\
    Position ($''$) & $x_s, y_s = (0, 0)$ \\
    \bottomrule
    \end{tabular}
    \vspace{-0.7cm}
    \end{center}
\end{table}

We compute time delay distances according to Equation (\ref{eq:D_dt}). The $H_0$ value, the source and the deflector redshifts are drawn from uniform prior distributions detailed in Table~\ref{tab:param}. We assume a flat $\Lambda$CDM cosmology. With the Fermat potential at the image positions and the time delay distance, we calculate the time delays from Equation (\ref{eq:time_delays}) and relative Fermat potentials from Equation (\ref{eq:Fermat}), meaning that doubles have one time delay - Fermat potential pair, while quads have three.

For the noise model, the goal is to emulate the results of a standard analysis, which models the system parameters from the lensing observation and measures the time delays from the image light curves. Therefore, we add Gaussian noise to the lensing parameters, the image positions and the source position. As standard deviations, we use each parameter's average error from the BNN in \citet{Park2021}. From those noisy estimates, we compute the Fermat potentials. For the time delays, we add Gaussian noise to the ones generated with the true parameters. This replicates the uncertainty yielded by the light curve measurements, as well as the mass-sheet degeneracy \citep{Park2021}. Table~\ref{tab:noise} summarizes all the standard deviations of the Gaussian noise distributions.

\begin{table}[t]
    \begin{center}
    \caption{Standard deviation of the Gaussian noise distributions used to mimic the uncertainties of lens modeling, time delay measurements, and image position measurements}
    \label{tab:noise}
    \begin{tabular}{lr}
        \toprule
        \textbf{Observables} & \textbf{Noise standard deviation} \\
        \midrule
        Time delays (days) & $0.35$\\
        Image positions ($''$) & $0.001$ \\
        \midrule
        \textbf{Deflector} & \\
        \midrule
        Position ($''$) & $0.005$ \\
        Einstein radius ($''$) & $0.011$\\
        Ellipticities & $0.039$ \\
        \midrule
        \textbf{External shear} & \\
        \midrule
        Components & $0.02$ \\
        \midrule
        \textbf{Active galactic nucleus} & \\
        \midrule
        Position ($''$) & $0.012$ \\
        \bottomrule
    \end{tabular}
    \end{center}
        \vspace{-0.8cm}
\end{table}


 \section{Methods}
\label{sec:NRE}

\subsection{Neural Ratio Estimation}
In this work, we train a Neural Ratio Estimator to learn the posterior distribution of $H_0$. At its core, a NRE learns the ratio between two distributions of the parameters of interest $\boldsymbol{\Theta}$ (in our case $H_0$), and the simulated observations $x$: the joint distribution $p \! \left(\mathbf{x}, \boldsymbol{\Theta}\right)$, which we can sample using our simulator, and the product of the marginals $p \! \left(\mathbf{x}\right) p \! \left( \boldsymbol{\Theta}\right)$, which we can sample by pairing randomly simulations and parameters sampled from the prior.  

Assigning the class label $y=1$ to the joint distribution and the class label $y=0$ to the product of the marginals, the optimal discriminator $\mathbf{d}^*$ that classifies samples from these two distributions converges to the decision function
\begin{equation}
    \mathbf{d}^* \! \left(\mathbf{x}, \boldsymbol{\Theta}\right) = p \! \left(y=1 \mid \mathbf{x}\right) = \frac{p \! \left(\mathbf{x}, \boldsymbol{\Theta}\right)}{p \! \left(\mathbf{x}, \boldsymbol{\Theta}\right) + p \! \left(\mathbf{x}\right) p \! \left( \boldsymbol{\Theta}\right)}
\label{eq:discriminator}
\end{equation}
The ratio $r \! \left(\mathbf{x} \mid \boldsymbol{\Theta}\right)$ between the distributions can be written as a function of the discriminator :
\begin{equation}
    r \! \left(\mathbf{x} \mid \boldsymbol{\Theta}\right) \equiv
    \frac{p \! \left(\mathbf{x}, \boldsymbol{\Theta}\right)}{p \! \left(\mathbf{x}\right) p \! \left( \boldsymbol{\Theta}\right)} =
    \frac{\mathbf{d}^* \! \left(\mathbf{x}, \boldsymbol{\Theta}\right)}{1 - \mathbf{d}^* \! \left(\mathbf{x}, \boldsymbol{\Theta}\right)}
\label{eq:ratio}
\end{equation}
The product between the estimator of $r$ learnt by the NRE, $\hat{r} \! \left(\mathbf{x} \mid \boldsymbol{\Theta}\right)$, and the prior distribution yields a posterior distribution estimator.
To conduct an inference with a trained Neural Ratio Estimator, the estimator $\hat{r}
\! \left(\mathbf{x} \mid \boldsymbol{\Theta}\right) $ is calculated multiple times for the same observation, but with different parameter values at each computation. 


\subsection{Set Transformer Architecture}

For the architecture of the discriminator, we use a Set Transformer \citep{Lee2019} to make use of the fact that different lensing configurations (doubles or quads) can have different number of time delay-relative Fermat potential pairs, and that those pairs are permutation invariant. We also explored Deep Sets \citep{Zaheer2017}, however in our experiments they were outperformed by the Set Transformer, and so we only report on the latter. 

The NRE takes as inputs the measured time delays, the modeled relative Fermat potentials, a $H_0$ value, the source's redshift, and the deflector's redshift. See Appendix~\ref{append:NN} Figure~\ref{fig:schema_discriminator} for the specific details of the architecture.

\subsection{Training}
\label{sec:train}
The training set, the validation set, and the test set contain 1,280,000 examples, 160,000 examples and 26,500 examples, respectively. The dataset is composed of approximately 83\% doubles and 17\% quads. We train the neural network on batches of 1,000 examples with a binary cross entropy loss as the objective function. At each batch, we draw a new realization of noise for the time delays, the parameters, the image positions, and the source position. We then compute the Fermat potentials. The training lasts for 5,000 epochs. The learning rate starts at $1 \times 10^{-4}$, and decreases by half every 500 epochs, as it was the optimal schedule we found through a hyperparameter search.


\begin{figure}[!t]
    \centering
    \includegraphics[width=.9\linewidth]{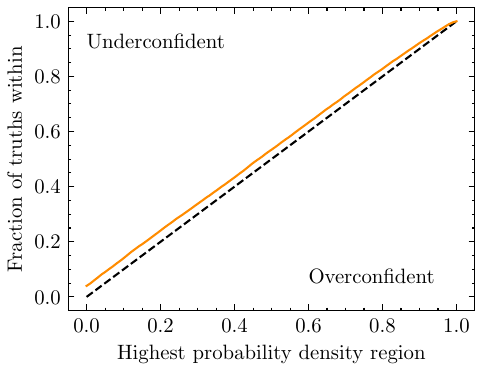}
    \caption{Coverage diagnostic of the NRE. A perfectly consistent distribution would fall on the dashed line. An underconfident distribution would lay on the top-left area, while an overconfident distribution would be in the bottom-right region. The NRE coverage, represented by the orange solid line, indicates a weak underconfident behaviour.}
    \label{fig:coverage}
    \vspace{-0.5cm}
\end{figure}

\section{Results and Discussion}
\label{sec:results}

In our framework, the general posterior in Equation (\ref{eq:general_posterior}) takes the specific form
\begin{equation}
    \begin{aligned}
    & P(H_0 | \Delta t, \Delta \phi, z_d, z_s) = \\
    & \int \, \mathrm{d}\boldsymbol{\zeta} \, \frac{P(\Delta t | H_0, \Delta \phi, z_d, z_s) P(\Delta \phi | \boldsymbol{\zeta}) P(\boldsymbol{\zeta}) P(H_0)}{P(\Delta t, \Delta \phi)}
     \label{eq:posterior}
\end{aligned}
\end{equation}
where $P(\Delta t | H_0, \Delta \phi)$ and $P(\boldsymbol{\zeta})$ are normal distributions, $P(\Delta \phi | \boldsymbol{\zeta})$ is a delta function, and $P(H_0)$ is a uniform distribution. 
We sample this posterior with \textsc{PolyChord} \citep{PolyChord1, PolyChord2} and find agreement with the NRE posteriors, as shown in some representative examples in Appendix \ref{append:posterios}.
To assess the NRE's accuracy, we perform a coverage test \citep{Hermans2021, Cole2022} using the highest posterior density (HPD) interval of the NRE on the noisy examples from the test set. Results are displayed in Figure \ref{fig:coverage}. The NRE shows a slightly underconfident behaviour, which is preferable to  overconfidence. 

Moreover, the NRE offers a significant improvement in the analysis speed. With \textsc{PolyChord}, the posterior sampling process requires from 20 to 40 minutes on a CPU, and is not amortized. By contrast, once trained, the NRE only requires $\sim$1 second to estimate the posterior of $H_0$ for a given lens, making the analysis more than 1000 times faster.

\begin{figure}[!t]
    \centering
    \includegraphics[width=.945\linewidth]{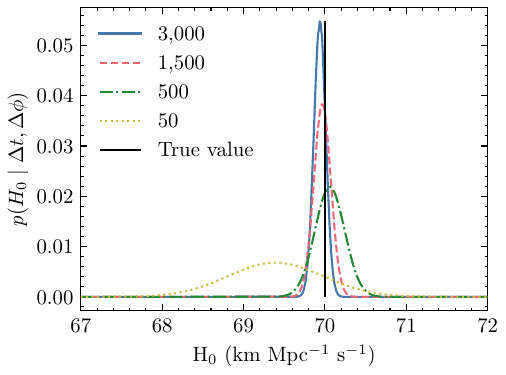}
    \vspace{-0.069cm}
    \caption{Population inferences of $H_0$ with the NRE. The blue solid line, the pink dashed line, the green dashed-dotted line, and the yellow dotted line represent populations of 3,000, 1,500, 500 and 50 lensing systems, respectively. The true value $H_0 = 70$ km s$^{-1}$ Mpc$^{-1}$ is indicated by the vertical black solid line. It falls inside the 2$\sigma$ interval for all populations.}
    \label{fig:inference}
    \vspace{-0.5cm}
\end{figure}

We perform a population inference of $H_0$. We simulate noisy data from multiple lensing systems (doubles and quads), fixing $H_0 = 70$ km s$^{-1}$ Mpc$^{-1}$. 
Figure \ref{fig:inference} shows the population inferences of 3,000, 1,500, 500 and 50 lensing systems. The NRE appears unbiased because all posteriors enclose the truth in their 2$\sigma$ interval.

One of the main advantages of a simulation-based approach such as the NRE over traditional maximum-likelihood methods is that it implicitly marginalizes over nuisance parameters \citep{Hermans2019}.  This is because, even though the simulator samples all parameters to generate the mock data, the classes and the loss function are independent of the nuisance parameters. While here our simulations remained simple, including further nuisance parameters in the inference is now reduced to simulating them.

Another important advantage of SBI methods is that they do not require any assumption about the form of the posterior. The complexity of the posterior is only limited by the simulations themselves, which can include complex environment, noise, selection effects, etc. In contrast, traditional explicit-likelihood methods require an analytical form for both the prior and the likelihood to compute the posterior distribution. These often imply simplistic priors, and simplifying assumptions about the model's parametrization, which can introduce biases in the inference.

A notable source of bias is the mass sheet degeneracy \citep{Falco1985}. In this paper, we do not consider explicitly the mass sheet degeneracy. However, we chose the noise distributions so that the uncertainty on $H_0$ could reach 8\% frequently, which is the error budget estimated by \cite{B&T2021} when accounting for the mass sheet degeneracy.






\section{Conclusion}
\label{sec:conclusion}

In this work, we used an NRE to infer $H_0$ from the time delays, the relative Fermat potentials, and the source and deflector redshifts of strong lensing systems. This work bridges the gap to completely amortize the inference of $H_0$ from time delay cosmography, bringing down the inference time by a factor of more than 1000 from at least 20 minutes with \textsc{PolyChord} to about 1 second per lens. Moreover, combining measurements from a population of 3,000 lenses suggests that our estimator is unbiased.

We assumed that the parameters describing the deflector could be estimated with a precision similar to that of BNNs published in the literature~\citep{Park2021}. To improve this work, more complex simulations incorporating environmental effects, such as the mass sheet degeneracy, as well as more inputs to break it, like velocity dispersion measurenments, could be used to train the NRE.

We expect the NRE to fully leverage the upcoming large datasets of strong lensing observations to reach the 1\% precision needed to solve the Hubble tension. Its implicit marginalization over nuisance parameters can take into account as many possible biases as can be simulated, while guaranteeing the accuracy of the inference. 

\section*{Acknowledgments}
This work was in part supported by Schmidt Futures, a philanthropic initiative founded by Eric and Wendy Schmidt as part of the Virtual Institute for Astrophysics (VIA). The work was in part supported by computational resources provided by Calcul Qu\'ebec and the Digital Research Alliance of Canada. Y.H. and L.P.L. acknowledge support from the Canada Research Chairs Program, the National Sciences and Engineering Council of Canada through grants RGPIN-2020-05073 and 05102, and the Fonds de recherche du Qu\'ebec through grants 2022-NC-301305 and 300397.

%




\bibliography{references}{}

\begin{thebibliography}{22}
\providecommand{\natexlab}[1]{#1}
\providecommand{\url}[1]{\texttt{#1}}
\expandafter\ifx\csname urlstyle\endcsname\relax
  \providecommand{\doi}[1]{doi: #1}\else
  \providecommand{\doi}{doi: \begingroup \urlstyle{rm}\Url}\fi

\bibitem[{Birrer} \& {Treu}(2021){Birrer} and {Treu}]{B&T2021}
{Birrer}, S. and {Treu}, T.
\newblock Tdcosmo - v. strategies for precise and accurate measurements of the
  hubble constant with strong lensing.
\newblock \emph{Astronomy \& Astrophysics}, 649:\penalty0 A61, 2021.
\newblock \doi{10.1051/0004-6361/202039179}.
\newblock URL \url{https://doi.org/10.1051/0004-6361/202039179}.

\bibitem[Cole et~al.(2022)Cole, Miller, Witte, Cai, Grootes, Nattino, and
  Weniger]{Cole2022}
Cole, A., Miller, B.~K., Witte, S.~J., Cai, M.~X., Grootes, M.~W., Nattino, F.,
  and Weniger, C.
\newblock Fast and credible likelihood-free cosmology with truncated marginal
  neural ratio estimation.
\newblock \emph{Journal of Cosmology and Astroparticle Physics}, 2022\penalty0
  (09):\penalty0 004, sep 2022.
\newblock \doi{10.1088/1475-7516/2022/09/004}.
\newblock URL \url{https://dx.doi.org/10.1088/1475-7516/2022/09/004}.

\bibitem[{Cranmer} et~al.(2015){Cranmer}, {Pavez}, and {Louppe}]{Cranmer2015}
{Cranmer}, K., {Pavez}, J., and {Louppe}, G.
\newblock {Approximating Likelihood Ratios with Calibrated Discriminative
  Classifiers}.
\newblock \emph{arXiv e-prints}, art. arXiv:1506.02169, June 2015.

\bibitem[{Falco} et~al.(1985){Falco}, {Gorenstein}, and {Shapiro}]{Falco1985}
{Falco}, E.~E., {Gorenstein}, M.~V., and {Shapiro}, I.~I.
\newblock {On model-dependent bounds on H(0) from gravitational images :
  application to Q 0957+561 A, B.}
\newblock \emph{Astrophysical Journal Letters}, 289:\penalty0 L1--L4, February
  1985.
\newblock \doi{10.1086/184422}.

\bibitem[Handley et~al.(2015{\natexlab{a}})Handley, Hobson, and
  Lasenby]{PolyChord1}
Handley, W.~J., Hobson, M.~P., and Lasenby, A.~N.
\newblock polychord: nested sampling for cosmology.
\newblock \emph{Monthly Notices of the Royal Astronomical Society: Letters},
  450\penalty0 (1):\penalty0 L61--L65, apr 2015{\natexlab{a}}.
\newblock \doi{10.1093/mnrasl/slv047}.
\newblock URL \url{https://doi.org/10.1093\%2Fmnrasl\%2Fslv047}.

\bibitem[Handley et~al.(2015{\natexlab{b}})Handley, Hobson, and
  Lasenby]{PolyChord2}
Handley, W.~J., Hobson, M.~P., and Lasenby, A.~N.
\newblock polychord: next-generation nested sampling.
\newblock \emph{Monthly Notices of the Royal Astronomical Society},
  453\penalty0 (4):\penalty0 4385--4399, sep 2015{\natexlab{b}}.
\newblock \doi{10.1093/mnras/stv1911}.
\newblock URL \url{https://doi.org/10.1093\%2Fmnras\%2Fstv1911}.

\bibitem[Hermans et~al.(2019)Hermans, Begy, and Louppe]{Hermans2019}
Hermans, J., Begy, V., and Louppe, G.
\newblock Likelihood-free mcmc with amortized approximate ratio estimators,
  2019.
\newblock URL \url{https://arxiv.org/abs/1903.04057}.

\bibitem[Hermans et~al.(2021)Hermans, Delaunoy, Rozet, Wehenkel, Begy, and
  Louppe]{Hermans2021}
Hermans, J., Delaunoy, A., Rozet, F., Wehenkel, A., Begy, V., and Louppe, G.
\newblock A trust crisis in simulation-based inference? your posterior
  approximations can be unfaithful, 2021.
\newblock URL \url{https://arxiv.org/abs/2110.06581}.

\bibitem[Hezaveh et~al.(2017)Hezaveh, Levasseur, and Marshall]{Hezaveh2017}
Hezaveh, Y.~D., Levasseur, L.~P., and Marshall, P.~J.
\newblock Fast automated analysis of strong gravitational lenses with
  convolutional neural networks.
\newblock \emph{Nature}, 548\penalty0 (7669):\penalty0 555--557, aug 2017.
\newblock \doi{10.1038/nature23463}.
\newblock URL \url{https://doi.org/10.1038\%2Fnature23463}.

\bibitem[{Kormann} et~al.(1994){Kormann}, {Schneider}, and
  {Bartelmann}]{Kormann1994}
{Kormann}, R., {Schneider}, P., and {Bartelmann}, M.
\newblock {Isothermal elliptical gravitational lens models.}
\newblock \emph{Astronomy \& Astrophysics}, 284:\penalty0 285--299, April 1994.

\bibitem[Lee et~al.(2019)Lee, Lee, Kim, Kosiorek, Choi, and Teh]{Lee2019}
Lee, J., Lee, Y., Kim, J., Kosiorek, A., Choi, S., and Teh, Y.~W.
\newblock Set transformer: A framework for attention-based
  permutation-invariant neural networks.
\newblock In Chaudhuri, K. and Salakhutdinov, R. (eds.), \emph{Proceedings of
  the 36th International Conference on Machine Learning}, volume~97 of
  \emph{Proceedings of Machine Learning Research}, pp.\  3744--3753. PMLR,
  09--15 Jun 2019.
\newblock URL \url{https://proceedings.mlr.press/v97/lee19d.html}.

\bibitem[Levasseur et~al.(2017)Levasseur, Hezaveh, and
  Wechsler]{PerreaultLevasseur2017}
Levasseur, L.~P., Hezaveh, Y.~D., and Wechsler, R.~H.
\newblock Uncertainties in parameters estimated with neural networks:
  Application to strong gravitational lensing.
\newblock \emph{The Astrophysical Journal}, 850\penalty0 (1):\penalty0 L7, nov
  2017.
\newblock \doi{10.3847/2041-8213/aa9704}.
\newblock URL \url{https://doi.org/10.3847\%2F2041-8213\%2Faa9704}.

\bibitem[Morningstar et~al.(2019)Morningstar, Levasseur, Hezaveh, Blandford,
  Marshall, Putzky, Rueter, Wechsler, and Welling]{Morningstar2019}
Morningstar, W.~R., Levasseur, L.~P., Hezaveh, Y.~D., Blandford, R., Marshall,
  P., Putzky, P., Rueter, T.~D., Wechsler, R., and Welling, M.
\newblock Data-driven reconstruction of gravitationally lensed galaxies using
  recurrent inference machines.
\newblock \emph{The Astrophysical Journal}, 883\penalty0 (1):\penalty0 14, sep
  2019.
\newblock \doi{10.3847/1538-4357/ab35d7}.
\newblock URL \url{https://dx.doi.org/10.3847/1538-4357/ab35d7}.

\bibitem[Oguri \& Marshall(2010)Oguri and Marshall]{O&M2010}
Oguri, M. and Marshall, P.~J.
\newblock Gravitationally lensed quasars and supernovae in future wide-field
  optical imaging surveys.
\newblock \emph{Monthly Notices of the Royal Astronomical Society}, pp.\
  no--no, apr 2010.
\newblock \doi{10.1111/j.1365-2966.2010.16639.x}.
\newblock URL \url{https://doi.org/10.1111\%2Fj.1365-2966.2010.16639.x}.

\bibitem[Park et~al.(2021)Park, Wagner-Carena, Birrer, Marshall, Lin, and
  Roodman]{Park2021}
Park, J.~W., Wagner-Carena, S., Birrer, S., Marshall, P.~J., Lin, J. Y.-Y., and
  Roodman, A.
\newblock Large-scale gravitational lens modeling with bayesian neural networks
  for accurate and precise inference of the hubble constant.
\newblock \emph{The Astrophysical Journal}, 910\penalty0 (1):\penalty0 39, mar
  2021.
\newblock \doi{10.3847/1538-4357/abdfc4}.
\newblock URL \url{https://doi.org/10.3847\%2F1538-4357\%2Fabdfc4}.

\bibitem[Pearson et~al.(2019)Pearson, Li, and Dye]{Person2019}
Pearson, J., Li, N., and Dye, S.
\newblock {The use of convolutional neural networks for modelling large
  optically-selected strong galaxy-lens samples}.
\newblock \emph{Monthly Notices of the Royal Astronomical Society},
  488\penalty0 (1):\penalty0 991--1004, 06 2019.
\newblock ISSN 0035-8711.
\newblock \doi{10.1093/mnras/stz1750}.
\newblock URL \url{https://doi.org/10.1093/mnras/stz1750}.

\bibitem[Riess et~al.(2022)Riess, Yuan, Macri, Scolnic, Brout, Casertano,
  Jones, Murakami, Anand, Breuval, Brink, Filippenko, Hoffmann, Jha,
  D’arcy~Kenworthy, Mackenty, Stahl, and Zheng]{Riess2022}
Riess, A.~G., Yuan, W., Macri, L.~M., Scolnic, D., Brout, D., Casertano, S.,
  Jones, D.~O., Murakami, Y., Anand, G.~S., Breuval, L., Brink, T.~G.,
  Filippenko, A.~V., Hoffmann, S., Jha, S.~W., D’arcy~Kenworthy, W.,
  Mackenty, J., Stahl, B.~E., and Zheng, W.
\newblock A {Comprehensive} {Measurement} of the {Local} {Value} of the
  {Hubble} {Constant} with 1 km s $^{-1}$ {Mpc} $^{-1}$ {Uncertainty} from the
  {Hubble} {Space} {Telescope} and the {SH0ES} {Team}.
\newblock \emph{The Astrophysical Journal Letters}, 934\penalty0 (1):\penalty0
  L7, July 2022.
\newblock ISSN 2041-8205, 2041-8213.
\newblock \doi{10.3847/2041-8213/ac5c5b}.
\newblock URL
  \url{https://iopscience.iop.org/article/10.3847/2041-8213/ac5c5b}.

\bibitem[{Schuldt} et~al.(2021){Schuldt}, {Suyu}, {Meinhardt}, {Leal-Taixé},
  {Ca\~nameras}, {Taubenberger}, and {Halkola}]{Schuldt2021}
{Schuldt}, S., {Suyu}, S.~H., {Meinhardt}, T., {Leal-Taixé}, L.,
  {Ca\~nameras}, R., {Taubenberger}, S., and {Halkola}, A.
\newblock Holismokes - iv. efficient mass modeling of strong lenses through
  deep learning.
\newblock \emph{Astronomy \& Astrophysics}, 646:\penalty0 A126, 2021.
\newblock \doi{10.1051/0004-6361/202039574}.
\newblock URL \url{https://doi.org/10.1051/0004-6361/202039574}.

\bibitem[Treu et~al.(2022)Treu, Suyu, and Marshall]{Treu2022}
Treu, T., Suyu, S.~H., and Marshall, P.~J.
\newblock Strong lensing time-delay cosmography in the 2020s.
\newblock \emph{The Astronomy and Astrophysics Review}, 30\penalty0
  (1):\penalty0 8, November 2022.
\newblock ISSN 1432-0754.
\newblock \doi{10.1007/s00159-022-00145-y}.
\newblock URL \url{https://doi.org/10.1007/s00159-022-00145-y}.

\bibitem[Wagner-Carena et~al.(2021)Wagner-Carena, Park, Birrer, Marshall,
  Roodman, Wechsler, and Collaboration)]{Wagner-Carena2021}
Wagner-Carena, S., Park, J.~W., Birrer, S., Marshall, P.~J., Roodman, A.,
  Wechsler, R.~H., and Collaboration), L. D. E.~S.
\newblock Hierarchical inference with bayesian neural networks: An application
  to strong gravitational lensing.
\newblock \emph{The Astrophysical Journal}, 909\penalty0 (2):\penalty0 187, mar
  2021.
\newblock \doi{10.3847/1538-4357/abdf59}.
\newblock URL \url{https://dx.doi.org/10.3847/1538-4357/abdf59}.

\bibitem[Weinberg et~al.(2013)Weinberg, Mortonson, Eisenstein, Hirata, Riess,
  and Rozo]{Weinberg2013}
Weinberg, D.~H., Mortonson, M.~J., Eisenstein, D.~J., Hirata, C., Riess, A.~G.,
  and Rozo, E.
\newblock Observational probes of cosmic acceleration.
\newblock \emph{Physics Reports}, 530\penalty0 (2):\penalty0 87--255, 2013.
\newblock ISSN 0370-1573.
\newblock \doi{https://doi.org/10.1016/j.physrep.2013.05.001}.
\newblock URL
  \url{https://www.sciencedirect.com/science/article/pii/S0370157313001592}.
\newblock Observational Probes of Cosmic Acceleration.

\bibitem[Zaheer et~al.(2017)Zaheer, Kottur, Ravanbakhsh, Poczos, Salakhutdinov,
  and Smola]{Zaheer2017}
Zaheer, M., Kottur, S., Ravanbakhsh, S., Poczos, B., Salakhutdinov, R., and
  Smola, A.
\newblock Deep sets, 2017.
\newblock URL \url{https://arxiv.org/abs/1703.06114}.

\end{thebibliography}
\bibliographystyle{icml2023}

\appendix

\section{Neural network variable dimensions}
\label{append:NN}

\begin{table*}[h]
    \caption{Input sizes for each operation in the deep set ratio estimator.}
    \label{tab:input_sizes}
    \vskip 0.15in
    \begin{center}
    \begin{tabular}{lc}
        \toprule
        \textbf{Operation} & \textbf{Input sizes} \\
        \midrule
        First multihead attention block & example set size $\times$ 2\\
        Second multihead attention block & example set size $\times$ 384\\
        Third multihead attention block & features : example set size $\times$ 384\\
         & learnable seed vector : example set size $\times$ 1 $\times$ 384\\
        Concatenating $H_0$ and the redshifts & 384 \\
        First linear layer & 387 \\
        Second linear layer & 768 \\
        Third linear layer & 768 \\
        Ratio estimation (see Equation (\ref{eq:ratio})) & 2 \\
        \bottomrule
    \end{tabular}
    \end{center}
\end{table*}

\begin{figure}[h!]
    \centering
    \includegraphics[width=1\linewidth]{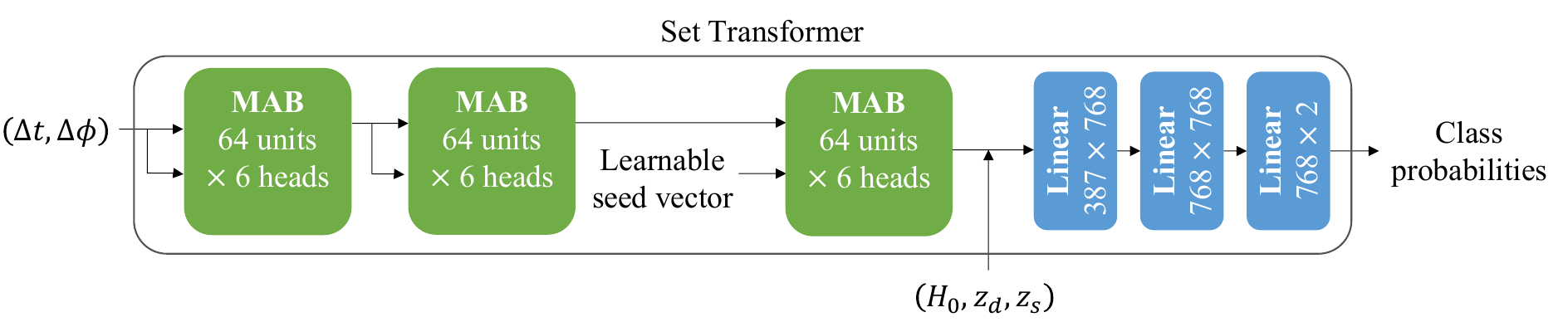}
    \caption{Architecture of the discriminator. The green squares represent the multihead attention blocks (MAB), and the blue rectangles, the linear layers. ELU activation functions are used after the first and the second linear layers. At inference, a softmax is added after the last layer.}
    \label{fig:schema_discriminator}
\end{figure}

Figure \ref{fig:schema_discriminator} and Table \ref{tab:input_sizes} illustrate our Set Transformer architecture. The first self-attention block computes multi-head attention between the time delay - relative Fermat potential pairs belonging to the same lensing system. The second self-attention block repeats the operation with the output of the first one. After, the features are aggregated by computing multi-head attention between a learnable seed vector and them. At each step, we use 6 attention heads of dimension 64. The $H_0$ value, $z_d$ and $z_s$ are concatenated to the result, which is then fed sequentially to 3 linear layers, each of 768 neurons. There is a ELU activation functions before and after the second layer. The whole neural network counts 2,224,514 parameters.

 At inference time, we apply a softmax function the final output to retrieve the class probabilities. We then insert the probability of the class with label $y=1$ in Equation (\ref{eq:ratio}) to estimate the distribution ratio. The latter is equivalent to the posterior density at the input $H_0$ because the prior is uniform.

\section{Examples of individual posteriors}
\label{append:posterios}

\begin{figure*}[h!]
  \centering
   \vspace{-0.7cm}
  \setlength{\tabcolsep}{0cm}
  \begin{tabular}[c]{ccc}
      & Quads & \\
      \includegraphics[height=3.9cm]{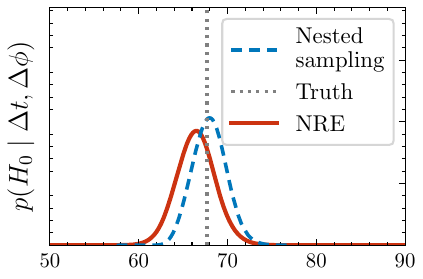}&
      \includegraphics[height=3.9cm]{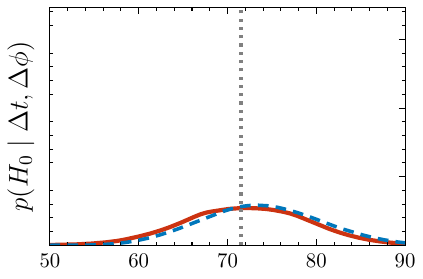}&
      \includegraphics[height=3.9cm]{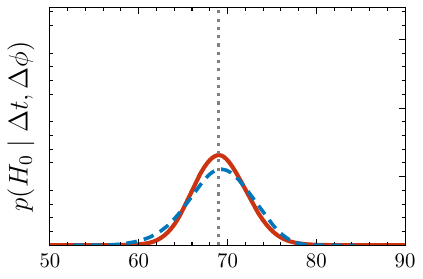}\\
      & Doubles & \\
      \includegraphics[height=3.9cm]{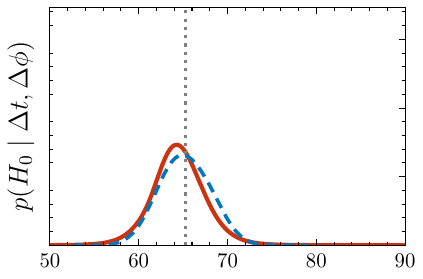}&
      \includegraphics[height=3.9cm]{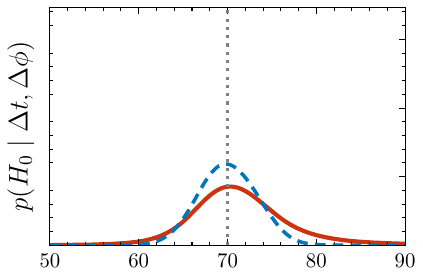}&
      \includegraphics[height=3.9cm]{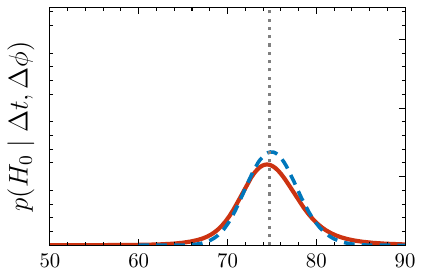}\\
       & $H_0$ (km Mpc$^{-1}$ s$^{-1}$) & \\
  \end{tabular}    
  \caption{Six $H_0$ inferences on examples from the test set. The first row show results for three different quads, and the last row, for three different doubles. The true $H_0$ value is also different on each plot. The blue dashed line indicates the true posterior distribution (computed with nested sampling and the true likelihood), the grey dotted line represents the true value, and the red solid line is the NRE posterior distribution. The difference between the two distributions is noticeable, but they still agree well with each other.}
  \label{fig:posteriors}
\end{figure*}

In Figure \ref{fig:posteriors}, we compare the NRE results on 6 representative test examples with those of nested sampling performed with the package \textsc{PolyChord} \citep{PolyChord1, PolyChord2}. Each plot is associated to a different lensing system and a different $H_0$ value. The nested sampling and the NRE posteriors are respectively indicated by the blue dashed line and the red solid line. The NRE shows a good agreement with the nested sampling posteriors. Moreover, each NRE posterior is a factor of about 1000 faster to obtain, taking only $\sim$1 sec on an NVidia V100 GPU, whereas sampling with \textsc{PolyChord} requires a minimum of 20 minutes per posterior.


\end{document}